\documentclass[a4paper]{spie}
\usepackage{upgreek}
\usepackage{url}
\usepackage{adjustbox}
\usepackage[normalem]{ulem}

\usepackage{amsmath,amsfonts,amssymb}
\usepackage{graphicx}
\usepackage[colorlinks=true, allcolors=blue]{hyperref}

\usepackage{float}
\usepackage{comment}
\usepackage{upgreek}

\newcommand{\todo}[1]{}
\renewcommand{\todo}[1]{{\color{red} TODO: {#1}}}

\title{ Characterizing and exploiting cross-talk effect in SPAD arrays for two-photon interference}

\author[a]{Sergei Kulkov}
\author[a]{Lada Radmacherova}
\author[b]{Ondrej Matousek}
\author[a]{Lou-Ann Pestana De Sousa}
\author[c]{Ermanno Bernasconi}
\author[c]{Claudio Bruschini}
\author[c]{Tommaso Milanese}
\author[c]{Edoardo Charbon}
\author[a,d,e]{Andrei Nomerotski}
\author[a,f]{Peter Svihra}

\affil[a]{Faculty of Nuclear Sciences and Physical Engineering, Czech Technical University in Prague,
	Břehová 7, Prague, Czech Republic}
\affil[b]{Faculty of Electrical Engineering, Czech Technical University in Prague,
	Technická 2, Prague, Czech Republic}
\affil[c]{Advanced Quantum Architecture Laboratory, École polytechnique fédérale de Lausanne (EPFL),
	Rue de la Maladière 71, Neuchâtel, Switzerland}
\affil[d]{Department of Electrical and Computer Engineering, Florida International University,
	10555 West Flagler St, Miami, U.S.A}
\affil[e]{Department of Astroparticle Physics, Institute of Physics of the Czech Academy of Sciences,
	Na Slovance 1999/2, Prague, Czech Republic}
\affil[f]{Department of Detector Development and Data Processing, Institute of Physics of the Czech Academy of Sciences, Na Slovance 1999/2, Prague, Czech Republic}
\authorinfo{Further author information: (Send correspondence to S. Kulkov)\\S. Kulkov: E-mail: sergei.kulkov@fjfi.cvut.cz}

\begin{document}
	\maketitle
	
	\begin{abstract}
		
		SPAD arrays are becoming a popular choice for measuring two-photon interference effects thanks to their high timing precision, fast readout, and high quantum efficiency. However, such sensors are affected by cross-talk that may mimic the useful signal. Even with a low probability of seeing cross-talk effect between neighboring SPADs, it was found that it still may reach the farther channels up to a half of millimeter away. Moreover, the use of microlenses that help SPADs achieve even better efficiency further boosts the cross-talk effect. In this work, we characterize the cross-talk effect and compare it to the Hanbury Brown-Twiss effect in the LinoSPAD2 camera, which has a linear sensor of 512 SPADs. Additionally, we compare the results between sensors with and without the microlenses. Finally, we present a timing calibration technique for the detector that utilizes the cross-talk effect.
	\end{abstract}
	
	
	\flushbottom
	
	\section{Introduction}
	\label{sec:intro}
	
\begin{figure}
		\centering
		\includegraphics[width=0.7\linewidth]{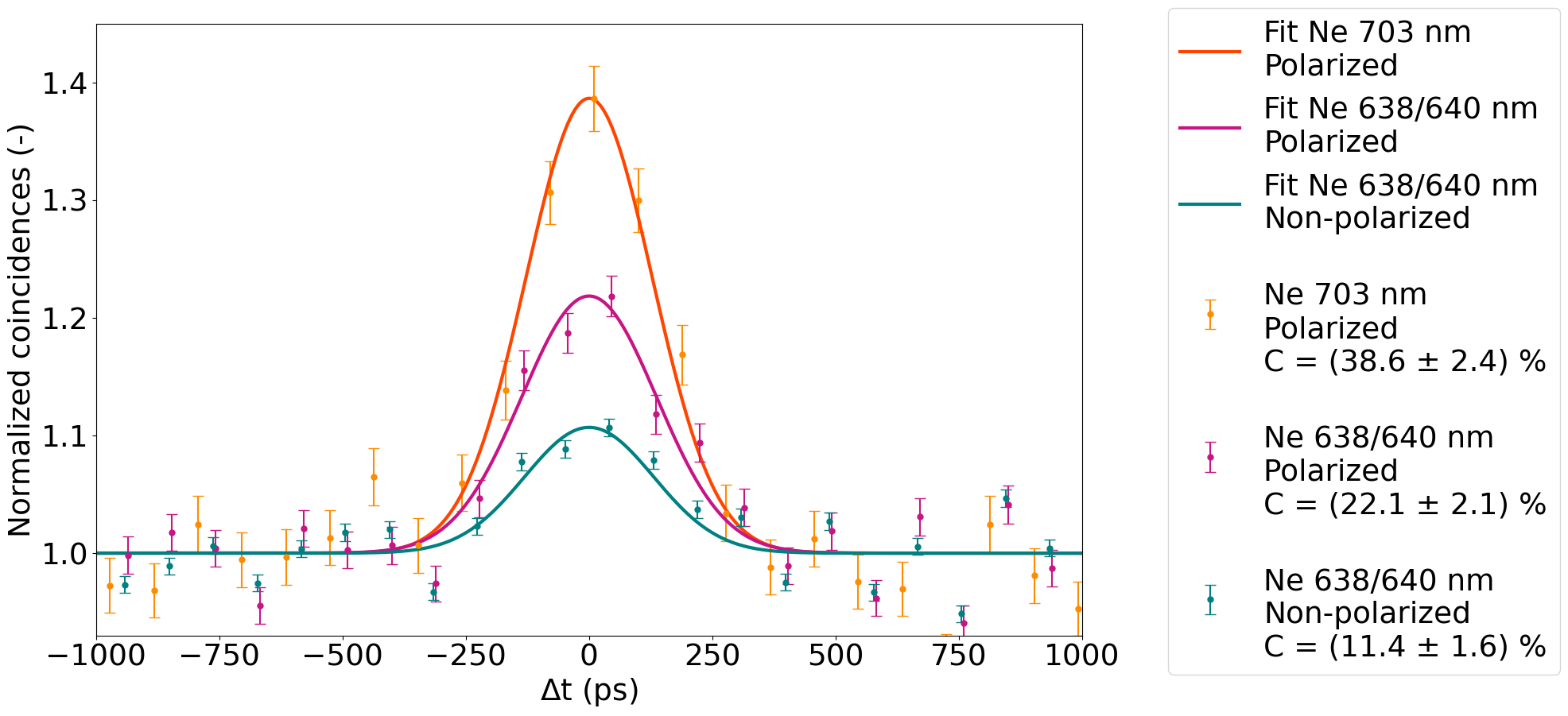}
		\caption{Contrast of the HBT effect depending on the photon wavelengths that reach the sensor. The HBT effect applies to photons of the same wavelength and the same polarization. In all three cases, an Ne lamp was used as a source of light with two different bandpass filters. The data are fitted using the Gaussian function to extract the contrast of the HBT peak. Three cases are showcased: first, with a polarizer and a bandpass filter ($700\pm5$) nm (orange); second, with a polarizer and filter ($640\pm5$) nm (violet); finally, no polarizer and filter ($640\pm5$) nm (teal). In the first case, the 703.2 nm spectral line is the strongest in the selected region of the Ne spectrum, resulting in almost all photons reaching the sensor being from that line. In the case of the ($640\pm5$) nm filter, there are two lines of approximately the same intensity: 638.3 and 640.2 nm, which leads to a mixing between the two different wavelengths, bringing the contrast of the peak down to half of the maximum possible. Additionally, if no polarizer is present, the contrast is halved due to the mix of two different polarizations detected by the sensor.}
		\label{fig:HBT_contrast}
	\end{figure}

	The Hanbury Brown and Twiss (HBT) effect for photons describes a tendency of thermal light to bunch in time, which arises from the super-Poissonian statistics that drive such sources \cite{loudon2000quantum, glauber1963quantum}. This correlation can be described with the second-order correlation function, which quantifies the probability that after detecting a photon at a certain point in time, other detections will follow within a short time interval. HBT and the Hong-Ou-Mandel effects \cite{HOM_effect, Bouchard_HOM_review, Sensors2020_Nomerotski} which are the two key two-photon interference effects are fundamental in such areas as quantum-assisted spectroscopy \cite{szoke2020entangled, mukamel2020roadmap, dorfman2021hong}, quantum imaging \cite{pittman1995optical, gatti2004correlated, erkmen2010ghost, lemos2014quantum, Farella2024}, and stellar intensity interferometry \cite{brown1956correlation, guerin2018spatial, rivet2020intensity, abeysekara2020demonstration, de2022combined, walter2023resolving, karl2024photon}.  A typical experimental setup designed for the observation of the HBT effect includes two independent single-photon sensitive detectors which timestamp each photon detected. From the collected data, the second-order correlation function $g^{(2)}(\vec{r}, \tau)$ is constructed and visualized in the form of a coincidence histogram. The normalized form of the function is defined for the source with intensity $I(\vec{r},t)$ as: 
	
	\begin{equation}
		g^{(2)}(\vec{r}, \tau) = \frac{ \textlangle I(\vec{r}, t) I(\vec{r}, t + \tau) \textrangle}{\textlangle I(\vec{r}, t) \textrangle ^ 2},
	\end{equation}
	where $\vec{r}$ is the distance between two detectors, $\tau$ is the time delay between two detections, and $\textlangle\textrangle$ denotes the averaging over time $t$. For thermal photons, this function will result in a peak in the coincidence histogram, i.e., an HBT peak. The maximum contrast for an HBT peak, i.e., the ratio of the peak height to the background level corresponding to Poissonian noise, is 1. In reality, the contrast measured is usually lower as the Gaussian response of the detector washes out the resolution, broadening the HBT peak and lowering the contrast. The HBT effect applies to photons of the same wavelength and polarization; in other words, the photons have to be indistinguishable. This can be seen in Fig. \ref{fig:HBT_contrast}, where three different situations are shown. When there are photons of multiple wavelengths reaching the sensor, e.g., from two separate spectral lines of the Ne spectrum, there are 4 possible pairs of detection events: two of the same wavelength and two of the mixed ones. Only the former two will contribute to the enhancement, i.e., the HBT effect, over the Possinoian noise. The same applies to polarization. Therefore, to measure the highest possible contrast of the HBT effect, either a narrow filter or high spectral resolution is required for selecting a spectral line or a narrow spectral bin, as well as a polarizer to achieve the same polarization of all photons reaching the sensor.
	
	There are many detector types capable of single-photon detection, e.g., time-stamping complementary metal-oxide-semiconductor cameras \cite{jachura2015shot, timepixcam, Nomerotski2019, Nomerotski2023}, charge-coupled device cameras \cite{jost1998spatial, brida2008measurement, zhang2009characterization, fickler2013real, avella2016absolute, reichert2017quality, moreau2019imaging}, transition-edge sensors \cite{cabrera1998detection, lita2008counting}, superconducting nanowire single-photon detectors \cite{divochiy2008superconducting,natarajan2012superconducting, holzman2019superconducting, zhu2020resolving, korzh2020demonstration}, and single-photon avalanche diodes (SPAD) \cite{charbon2014single, perenzoni2016compact, gasparini2017supertwin, bruschini2019single, morimoto2020megapixel, lubin2021heralded, wojtkiewicz2024review}. Sensors based on the latter have recently become a very popular option in physics applications, thanks to their rapid development. Operational benefits include high photon detection efficiency (PDE), defined as the ratio of the number of detected photons to the number of incident photons, and the fact that the signals and readout are fully digital. The digital form of the signal is generally more stable than the analog form, and the processing of the signal is easier and faster. The sensors are also compact and have a reasonable cost, which makes them accessible. In addition, silicon SPADs can be operated at room temperature which makes them highly versatile.
	
	However, SPAD sensors also have drawbacks. Due to a charge-multiplication mechanism and their small pixel pitch, they are particularly susceptible to afterpulsing \cite{ziarkash2018comparative} and cross-talk (CT) \cite{rech2008optical}. There are two processes that cause CT: the generation of secondary photons, referred to as optical cross-talk, and capacitive coupling, which is the transfer of energy between different segments of a circuit, known as electrical cross-talk \cite{rech2007depth, rech2008optical, xu2014crosstalk, xu2016cross, ficorella2016crosstalk, kroger2017high, ficorella2017crosstalk, jahromi2018timing, ratti2021cross}. By proper engineering, the electrical cross-talk can be reduced to a negligible level, leaving the optical CT as the primary contributor. When a photon from a light source (signal photon) is incident on the active region of a SPAD, an electron avalanche is formed. During the avalanche, a secondary photon can be emitted (CT photon), which can travel to adjacent pixels of the sensor and trigger another avalanche. The new avalanche is registered as a separate event, leading to spurious detections.
	
	\begin{figure}[ht]
		\centering
		\includegraphics[width=0.7\linewidth]{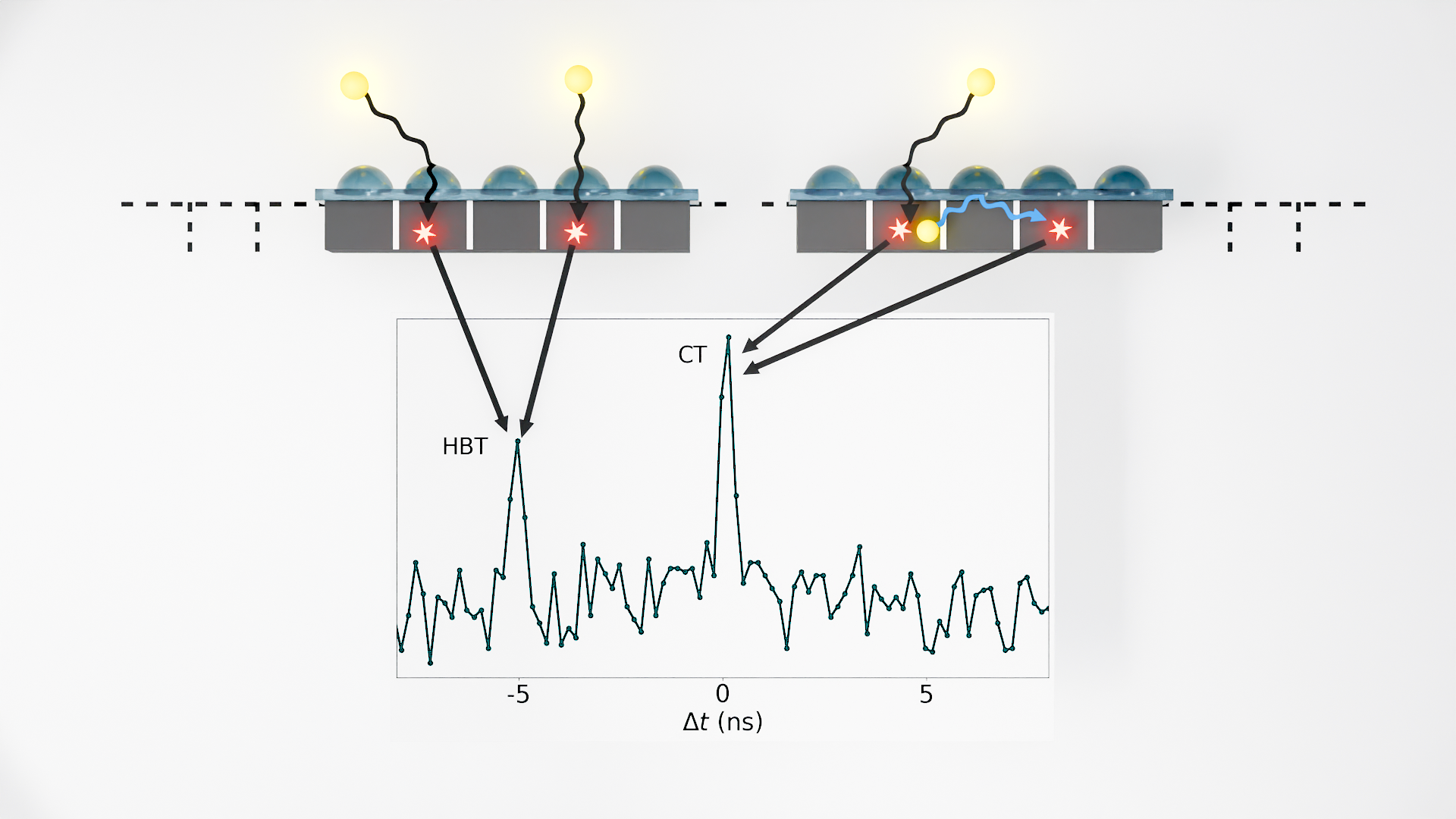}
		\caption{A schematic depiction of the HBT effect (left) and the CT effect (right) in a silicon SPAD array with microlenses. Both effects rely on the detection of two photons. In the case of the HBT effect, both photons come from a source of light, while only one signal photon is required for the CT. The second photon that contributes to the CT effect is generated in the sensor itself during an electron avalanche that follows a detection of a signal photon. That secondary photon may reach neighboring pixels, triggering another electron avalanche and resulting in a parasitic signal. Microlenses deposited above each SPAD help improve the PDE of the sensor by guiding the incoming photons onto the photosensitive area. As a drawback, they can act as a reflective layer, sending CT photons back to the sensor, which, as a result, may travel farther, reaching more distant channels. The HBT peak can be shifted relative to the CT one by delaying one of the signal photons before they reach the sensor.}
		\label{fig:CT_in_ML}
	\end{figure}
	
	This becomes a serious issue, particularly in two-photon coincidence measurements, such as the measurement of the HBT effect, see Fig. \ref{fig:CT_in_ML}. To observe the effect, two photons have to arrive at two different detectors at a similar time. For CT, on the other hand, only one photon is required, where the second is generated in the sensor itself. When the secondary, CT photon is detected, the HBT effect is mimicked \cite{kulkov2024inter}. The probability of CT decreases with the distance between the pixel where the primary photon is registered and the pixel where the CT photon generates another avalanche. In the coincidence histogram, both effects result in a peak at the same position near $\Delta \textit{t}$ = 0. However, CT is inherent to the detector, and the position of the CT peak depends only on the delays added during signal propagation from the pixel to the timestamping modules. In contrast, the position of the HBT peak can be adjusted by extending the path of one of the photons, separating the CT and the HBT peaks as a result.
	
	Linear arrays of SPADs open the way to new applications impossible with single-channel devices, e.g., fast spectrometers near the Heisenberg Uncertainty Principle \cite{jirsa2025fast, ferrantini2025multifrequency}. Therefore, it is necessary to study and characterize the CT effect in such devices. In this work, we characterize the dark count rate (DCR) and the CT effect of the LinoSPAD2 detector equipped with a microlensed sensor, comparing our findings with those from a detector without microlenses. Additionally, we simultaneously measure and study the CT effect and how it compares to and influences the HBT effect. Finally, we investigated how the CT effect can be used for calibrating the system.
	
	\section{Methods}
	\subsection{The LinoSPAD2 detector}
	
	In this work, we used the LinoSPAD2 camera \cite{milanese2023linospad2, bruschini2023linospad2}, which is a single-photon detector with a linear sensor consisting of 512 SPADs with a pixel size of 26.2 $\times $ 26.2 $\upmu\mathrm{m}^{2}$. Each photon registered is timestamped via time-to-digital converters (TDC) inside one of the two Spartan 6 field-programmable gate arrays (FPGA), where each FPGA reads out half of the sensor. In this work, we utilized only one half of the sensor. Besides the timestamp itself, the data output also includes information about the pixel where the detection occurred, so it is possible to reconstruct the sensor occupation. The timing precision of the detector is approximately 40 ps rms \cite{jirsa2025fast}, the maximum PDE is 28 \% at 520 nm wavelength, and the fill factor is 25.1\%. The fill factor can be further increased by a factor of 2.3 to approximately 58\% with the use of microlenses that are placed on top of each SPAD and direct the light onto the photosensitive area \cite{bruschini2023challenges}. In the previous works \cite{milanese2023linospad2, bruschini2023linospad2}, the DCR was estimated at approximately 70 cps/pixel and the average cross-talk probability is 0.2\% for the immediate neighboring pixels, both measured at room temperature and applied bias voltage of 28 V, where 24 V is the breakdown voltage of the sensor\cite{milanese2023linospad2}. These operational parameters allow the detector to operate under ambient conditions without the need for cooling to extremely low temperatures.
	
	Each TDC consists of 35-element delay lines made of 4-bit carry chain blocks. The total number of timing bins in each TDC is, therefore, 140. The measurement is synchronized to the 400 MHz clock provided by a crystal oscillator located on the motherboard, resulting in the average TDC bin width of 17.857 ps. The real widths of these bins are different, and to account for that, the system is calibrated to compensate for these TDC nonlinearities by calculating a look-up table \cite{SK_dissertation}. Additionally, the system can be calibrated to compensate for differences in electrical paths from the individual SPADs to the FPGA, which, ideally, can be done with a very fast laser \cite{SK_dissertation}. In this work, however, we explored how CT can be harnessed to calculate and calibrate these offsets.
	
	\subsection{Experimental setup}
	
	\subsubsection{Characterizing dark count rate and cross-talk}
	
	\begin{figure}[ht]
		\centering 
		\adjustbox{valign=c}{\includegraphics[width=.6\textwidth]{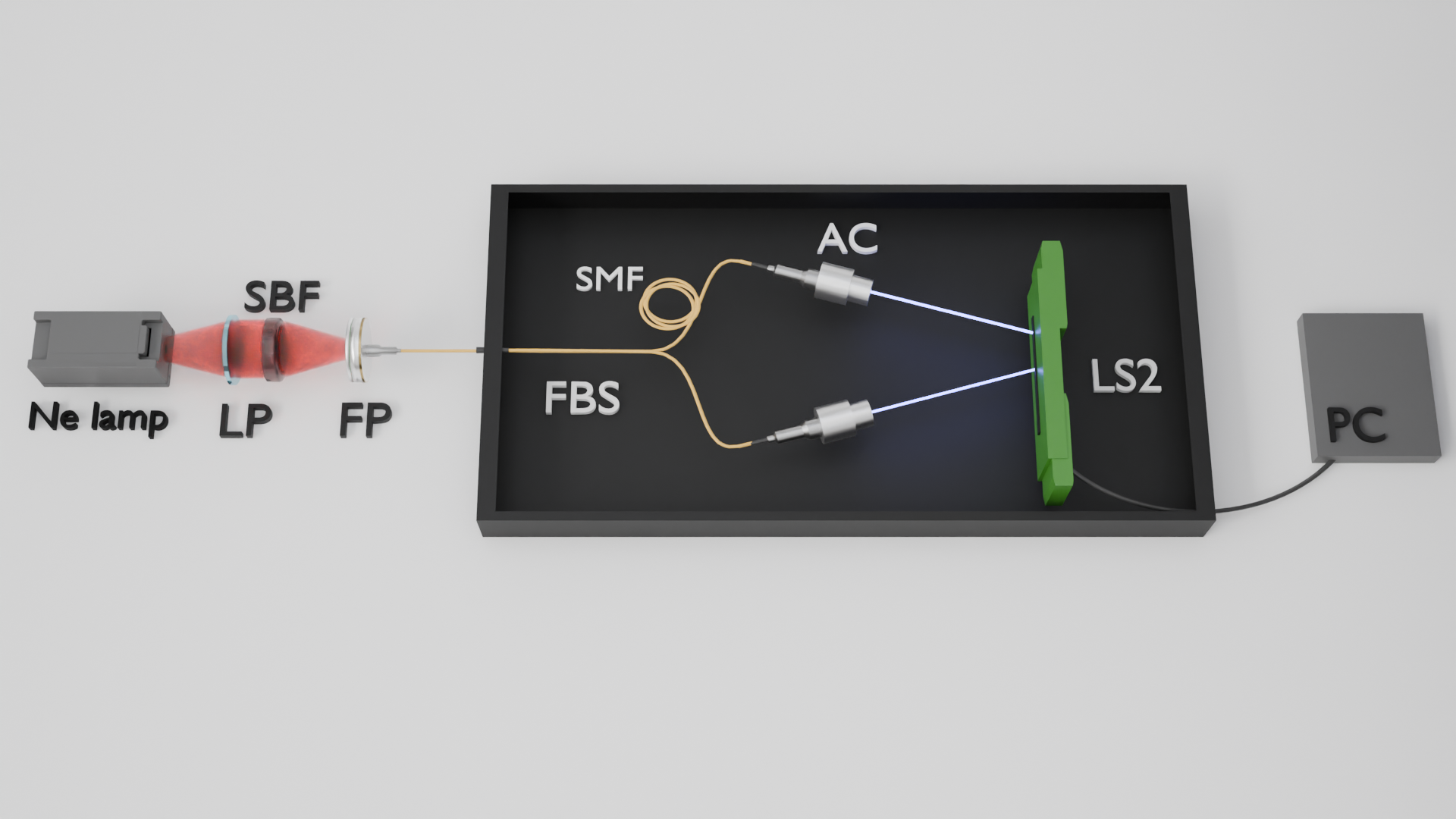}}
		\qquad
		\adjustbox{valign=c}{\includegraphics[width=.335\textwidth]{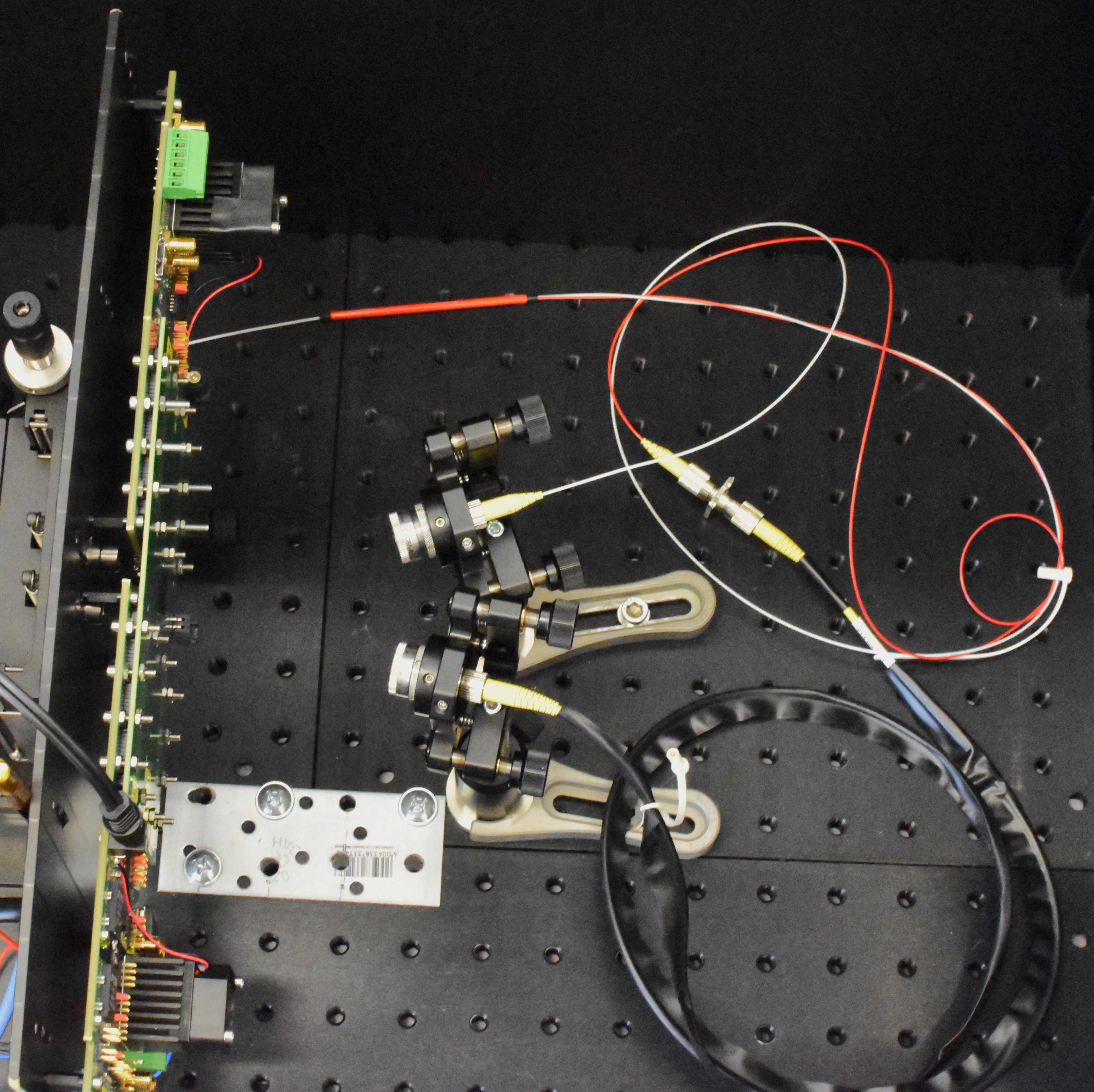}}
		\caption{\label{fig:setup} Left: schematic of the setup. Light from the Ne lamp is linearly polarized (LP) and spectrally filtered by a spectral band-pass filter (SBF) and collected into a 1-to-2 50:50 single-mode fiber beamsplitter (FBS) through a fiber port (FP). An additional delay is introduced into the path of one of the photons in the form of a 1-meter-long single-mode optical fiber (SMF) to delay one of the two HBT photons. The outputs of the beamsplitter are directed into two adjustable collimators (AC) that focus the beams onto the LinoSPAD2 sensor (LS2). The data from the detector are transferred to the PC for analysis. Right: photo of the LinoSPAD2 camera and the two adjustable collimators with the fiber beamsplitter inside the light-tight enclosure.}
	\end{figure}
	
	First, to characterize the board in ambient conditions and without any optical signal, we measured the DCR in conditions that replicate our two-photon interference measurements. Figure \ref{fig:setup} shows a schematic of the usual experimental arrangement that we use for measuring the HBT effect and a photo of the detector. During the DCR measurements, the light sources were powered off. Due to the single-photon sensitivity, the sensor has to be covered from ambient light, and to achieve that, we placed the detector inside a light-tight enclosure and covered any sources of light, e.g., light-emitting diodes (LED) present on the motherboards. There is a cooler on top of each of the two FPGAs, but since the whole board is enclosed in a relatively small space, the temperature does rise inside with time. To see how much and how fast does the temperature change and how it affects the DCR and CT numbers, we taped a thermometer probe next to the sensor. The data from the probe were sent to a PC via a USB3 connection, and the temperature measured was saved every 10 s for approximately 4 hours. The room temperature in the lab measured outside the black box was at $24^{\circ}\mathrm{C}$. The data from the detector were sent to a PC via a USB3 connection.
	
\begin{figure}[ht]
		\centering 
		\adjustbox{valign=c}{\includegraphics[width=.47\textwidth]{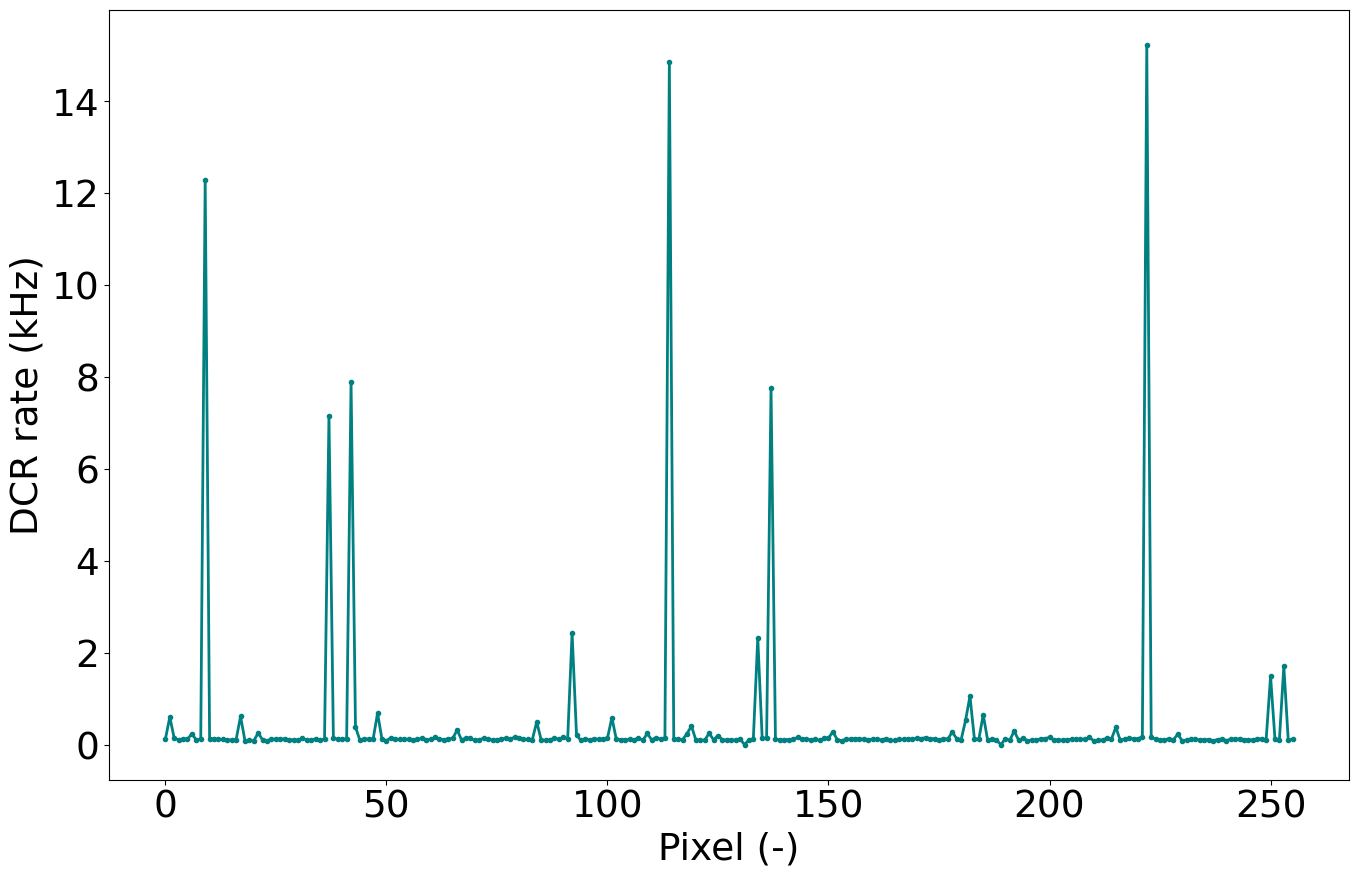}}
		\qquad
		\adjustbox{valign=c}{\includegraphics[width=.47\textwidth]{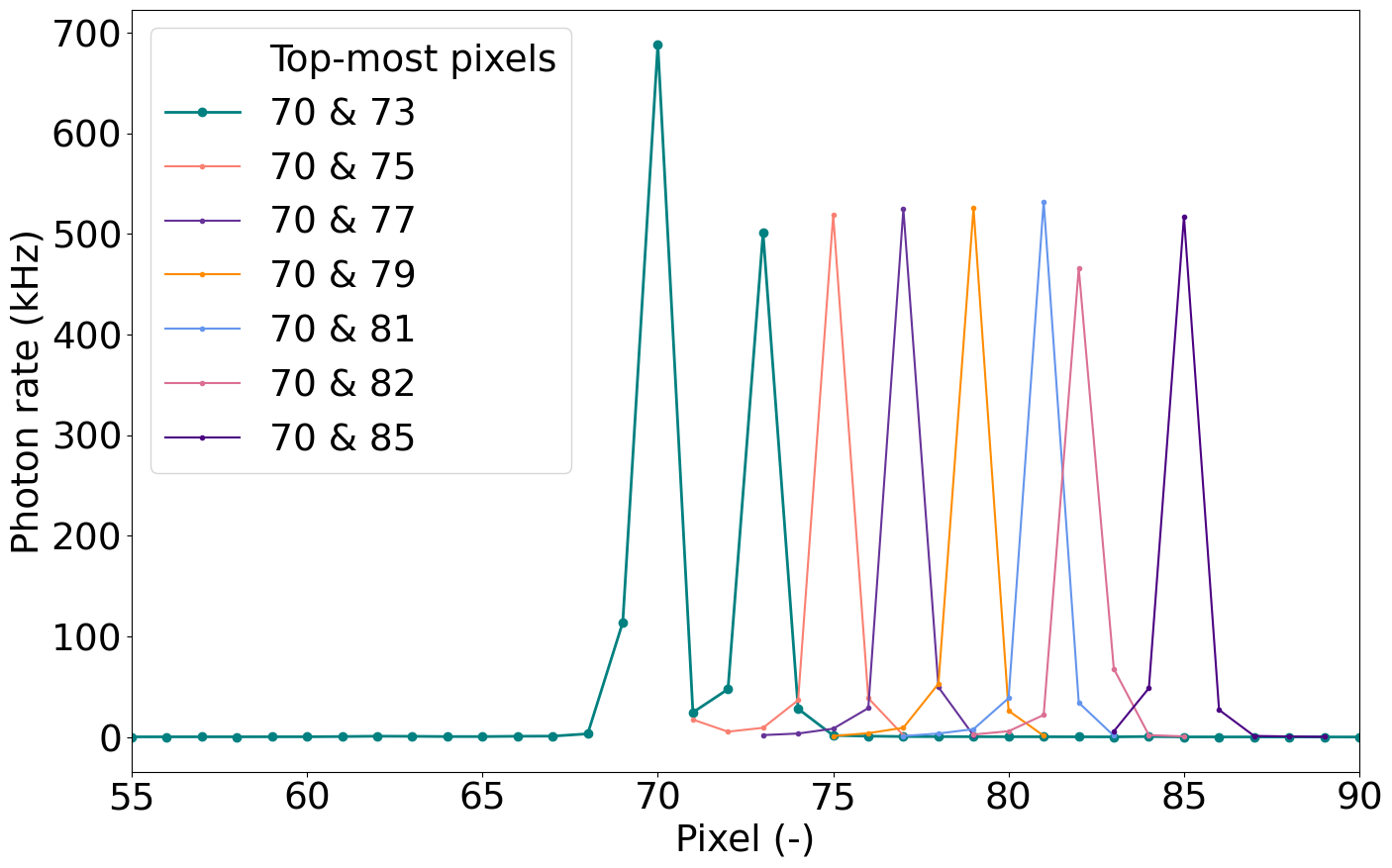}}
		\caption{\label{fig:senpop} Left: Sensor population plot showing only the DCR. There are 11 hot pixels with a DCR above 1 kHz. Right: Sensor population plot for 7 data sets, where one beam of Ne light was always focused at pixel 70 and the other one was moved further away from it with each consequent measurement. The pair of pixels where the two beams are focused are shown for each data set in the legend.}
	\end{figure}

	Additionally, using the same setup --- and the same data --- one can calculate the average CT probability and its dependency on the distance between the two pixels of interest using hot pixels. Hot pixels are noisy SPADs that generate a lot of signals, sometimes reaching kHz. For the sensor half used in this work, there were measured 11 hot pixels with DCR exceeding 1 kHz, see the left panel in Fig. \ref{fig:senpop}. These contribute to the CT measured across the whole sensor, affecting mostly neighbors up to 20 pixels away. Looking for photon coincidences between two different pixels where one of them is a hot pixel will likely --- depending on how far the two pixels are --- produce a CT peak, which can be used to measure the CT probability at the given distance.
	
	There are two available configurations of the LinoSPAD2 camera: with and without the microlenses on top of the sensor. In our previous work \cite{kulkov2024inter}, we characterized the CT and HBT effects in the LinoSPAD2 without the microlenses. This time, however, we had a camera with microlenses, and we were able to compare both the DCR and CT effects between the two configurations. DCR is not expected to change much, but CT may be affected by structures present on top of the sensor since the secondary photon can be reflected off these structures back into the sensor \cite{Nakamura2019}.
	
	\subsubsection{Exploiting cross-talk for offset calibration}
	
	In LinoSPAD2, photon timestamping is performed by the FPGA, which resides on a separate circuit board from the sensor. Electrical paths between the individual pixels and the FPGA differ in length, and this results in intrinsic delays being added to all timestamps. This results in an offset of the two-photon coincidence effects, shifting both the CT and HBT peaks in the coincidence histograms. To compensate for that, one can estimate these individual delays and subtract them post-factum, shifting the peaks to $\Delta t=0$.
	
	This could be done by shining a fast laser upon the whole sensor. This allows one to isolate the delays as the timestamps of the laser photons would be essentially the same. In this work, we propose a faster and cheaper alternative where the CT effect is utilized instead.
	
	First, we collect data under ambient light or by using a weak light source (e.g., an LED) shining upon the whole sensor. Then, we compute the photon coincidence histograms, where the CT effect will produce CT peaks. Here, we take advantage of the fact that the CT effect is the strongest for the adjacent SPADs. We fit each peak using the Gaussian function to extract its position $\Delta t \neq 0$. Let $t_i$ and $t_j$ be the timestamps of two coinciding photons produced by the CT effect in the $i$-th and $j$-th pixels, where $i \in \{0, \ldots, 254\}$ and $j \in \{1, \ldots, 255\}$. The offset $\mathrm{off}_{i,j}$ for the two pixels that correspond to the measured position of the CT peak can be written as:
	\begin{equation}
		(t_i + d_i) - (t_j + d_j) = d_i - d_j = off_{i,j}, \quad i \in \{0, 254 \}, \quad j \in \{1, 255 \},
	\end{equation}
	where $d_i$ and $d_j$ are the delays introduced to each pixel's timestamp. The offsets measured in LinoSPAD2 range from hundreds of ps up to $\sim10$ ns. Compared to that, the timestamp difference between the signal and the CT photon can be estimated on the order of ps, given by the small pixel pitch of 26.2 $\upmu\mathrm{m}$. Therefore, we estimate that the two timestamps corresponding to the CT peak are essentially equal: $t_i = t_j$. We enclose the system of equations with a requirement for the average delay equaling zero, corresponding to the ideal case where no delays and, hence, no offsets are introduced:
	\begin{equation}
		\frac{1}{256}\sum_{k=0}^{255} d_k = 0.
	\end{equation}
	
	The whole offset calibration problem can be formulated together in matrix form using a bidiagonal matrix $\mathbb{A}$ with a nonzero last row as follows:
	\begin{equation}
		\mathbf{\mathbb{A}x=b},
		\label{eq:linsys}
	\end{equation}
	where
	\begin{equation}
		\mathbb{A} = 
		\begin{bmatrix}
			1 & -1 & 0 & 0 & \cdots & 0 & 0 \\
			0 & 1 & -1 & 0 & \cdots & 0 & 0 \\
			0 & 0 & 1 & -1 & \cdots & 0 & 0 \\
			\vdots & \vdots & \vdots & \vdots & \vdots & \ddots & \vdots \\
			0 & 0 & 0 & 0 & \cdots & 1 & -1 \\
			1 & 1 & 1 & 1 & \cdots & 1 & 1
		\end{bmatrix}, \space \mathbf{x} = \begin{bmatrix} d_0 \\ d_1 \\ d_2 \\ \vdots \\ d_{254} \\ d_{255} \end{bmatrix}, \space \mathbf{b} = \begin{bmatrix} off_{0,1} \\ off_{1,2} \\ off_{2,3} \\ \vdots \\ off_{244,245} \\ 0 \end{bmatrix}.
	\end{equation}
	Solving this system of equations, one will get a vector of 256 delays that can be used during data analysis for offset corrections. We use the offset calibration calculated this way for all results presented in this work.
	
	The procedure described above can be applied for the LinoSPAD2 detector with firmware version "2212s". For the firmware version "2212b" see Ref. \citenum{SK_dissertation}. The two procedures are somewhat different due to the different ways individual SPADs are connected to the corresponding TDCs.
	
	\subsubsection{Joint cross-talk and HBT measurements}
	
	For the purpose of the joint HBT and CT measurements, we used a Ne lamp, collecting light into single-mode optical fibers and focusing two beams onto two different pixels. The experimental arrangement used for these measurements can be seen in Fig. \ref{fig:setup}. The Ne lamp is powered by an AC source at 240 V, 10 mA, and 50 Hz. The light from the lamp is collected via a fiber port (Thorlabs PAF2-5B) which also holds a spectral filter (Thorlabs FBH640-10) with central wavelength of 640 nm and full width at half maximum of 10 nm used to select the 638/640 nm Ne doublet, and a polarizer (Thorlabs LPNIRE100-B) which is used to select a single polarization to achieve a two-fold increase in HBT peak contrast. Through the fiber port, the light continues into a 1-to-2 50:50 fiber beamsplitter (Thorlabs TW670R5F1), which divides the light into two beams. The two outputs of the beamsplitter are connected to adjustable collimators (Thorlabs CFC8A-A) used to focus the two beams onto different SPADs. Each collimator is mounted on an optics holder (Thorlabs KM100) that allows to steer each beam freely, covering the 2D plane where the sensor is located. The measurements were performed for seven different pairs of pixels with gradually increasing distances between them; one beam was always focused at pixel 70, see the right panel in Fig. \ref{fig:senpop}. Besides the PC and past the fiber port, everything is placed inside the light-tight enclosure to avoid light contamination of the useful signal and prevent the sensor from collecting noise.
	
	For each acquisition cycle, timestamp differences $\Delta \textit{t}$ were calculated. The timestamp difference represents the length of a time interval between registering a photon at one pixel and a subsequent detection in another. The data collection duration was less than 45 minutes for each of the seven pairs of pixels. CT can interfere with measurements of the HBT effect due to false coincidences generated by CT photons, which produce a peak near the $\Delta \textit{t}$ = 0 position in the coincidence histogram. In an unadjusted setup, the HBT peak will appear at the same position, and, as a result, the two peaks will overlap. To distinguish between the two, we insert an additional 1 m long optical fiber into one of the beamsplitter's arms, delaying the arrival time of the photons traveling through the prolonged arm by approximately 5 ns. This leads to a shift in the position of the HBT peak and prevents an overlap with the CT one.
	
	\section{Results and discussion}
	
	\subsection{Dark count rate and cross-talk measurements}
	
	\begin{figure}[ht]
		\centering 
		\adjustbox{valign=c}{\includegraphics[width=.47\textwidth]{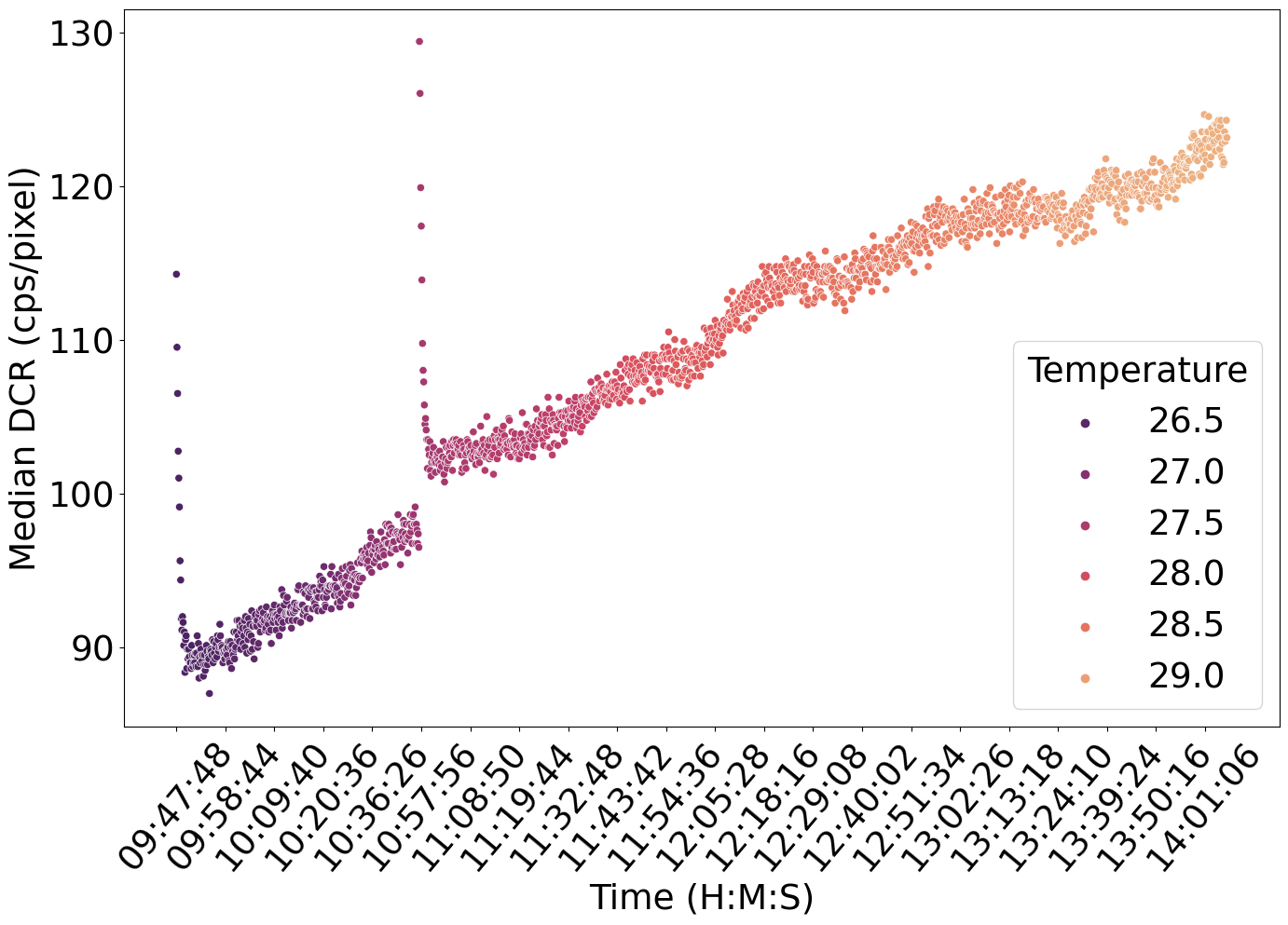}}
		\qquad
		\adjustbox{valign=c}{\includegraphics[width=.47\textwidth]{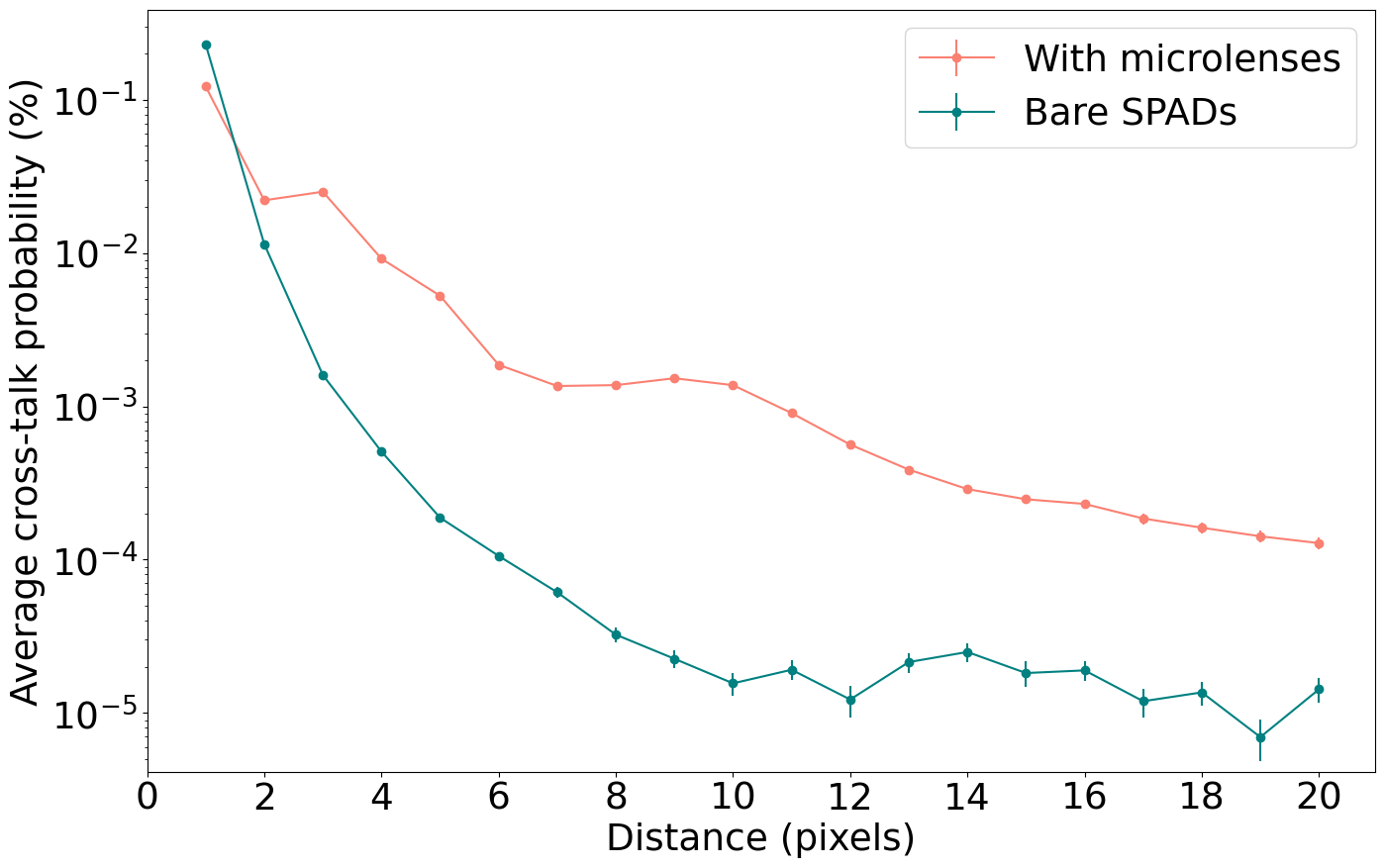}}
		\caption{\label{fig:DCR_CT} Left: Evolution of the median DCR in time and with rising temperature. The average temperature measured next to the sensor is shown in the legend, with each number corresponding to one of the 6 equal subsets measured continuously. The measurement was conducted for over 4 hours, through the course of which the temperature rose by $2.5^{\circ}\mathrm{C}$. The median DCR was measured at 107 cps/pixel, and at the end of the measurement, an increase of $\sim30\%$ was measured. Right: Dependency of the average CT probability on the distance between the two pixels of interest, measured for the sensor with microlenses (in red) and compared to the previously done measurement for the sensor without \cite{kulkov2024inter}. The CT probability for the immediate neighbor was estimated at 0.12\% for the sensor with microlenses --- lower than the 0.19\% probability for the sensor without. The CT effect is much stronger in the presence of microlenses, leading to enhanced CT at a distance of 3 pixels and a constant probability for the distances from 6 to 11 pixels.}
	\end{figure}
	
	To measure the median DCR and how it changes with time and temperature, we collected data for approximately 4 hours straight. The resulting dataset was divided into 6 subsets, where the median DCR and the average temperature measured were calculated for each. The result can be seen in the left panel in Fig. \ref{fig:DCR_CT}. In the beginning of the measurements, the DCR was estimated at $\sim90$ cps/pixel. Nearing the end of the measurement, the temperature rose for $2.5^{\circ}\mathrm{C}$. As a result, the DCR rose 30\% up to 120 cps/pixel. The median DCR over the whole dataset was estimated at 107 cps/pixel --- which is a comparable number to our previous measurements with LinoSPAD2 \cite{kulkov2024inter}, though a bit higher than what is reported in the original work \cite{milanese2023linospad2}. This can be attributed to the fact that our setup is far from ideal for such measurements, and the setup we use for DCR and CT measurements resembles the one we use when working with thermal light. Additionally, some ambient light can still reach the sensor both from outside and the onboard LEDs due to the nonideal shielding of the sensor from such light. The two occasional spikes in the DCR level can be addressed to manipulations near the setup, which resulted in ambient light reaching the sensor. Even though the 30\% increase may seem significant, the data throughput covered by DCR still takes only a few percent even at higher temperatures, and the DCR level is way below the photon rates we achieve with light sources, which is typically above tens of kHz and reaching MHz for some light sources. For longer data collection, the DCR is expected to stabilize at a level of approximately 125 cps/pixel \cite{kulkov2024inter}.
	
	The average CT probability and its dependency on the distance between the two pixels firing was calculated using 8 hot pixels with the highest DCR and their neighbors on both sides where applicable and up to 20 pixels away. The probability was also averaged over the 6 data sets measured at different temperatures, same as DCR. For each distance in pixels, the CT probability is an average over 14 pairs of pixels. The result can be seen in the right panel in Fig. \ref{fig:DCR_CT}. For the immediate neighbor, the CT probability was estimated at 0.12\% for the sensor with microlenses, compared to 0.19\% we measured for the sensor without microlenses in the previous work \cite{kulkov2024inter}. Furthermore, for farther neighboring pixels, the CT probability is much higher in the presence of microlenses, especially at a distance of 3 pixels. This may be attributed to the additional layer of material above each SPAD, which may lead to reflections of the secondary, CT photons back to the sensor, with seemingly the strongest reflection for the third neighbor. Additionally, a plateau for the distances from 6 to 11 pixels is present for the sensor with microlenses. This fact has more severe consequences for such sensors as the CT does not disappear as quickly with distance as in the sensors without the microlenses, further complicating measurements with two-photon interferences. Interestingly, no effect of rising temperatures near the sensor on the CT probability was measured, which is expected since the CT is primarily dependent on the material properties and sensor and digital readout architecture.
	
	\subsection{Joint cross-talk and HBT measurements}
	
	\begin{figure}[ht]
		\centering
		\includegraphics[width=0.7\linewidth]{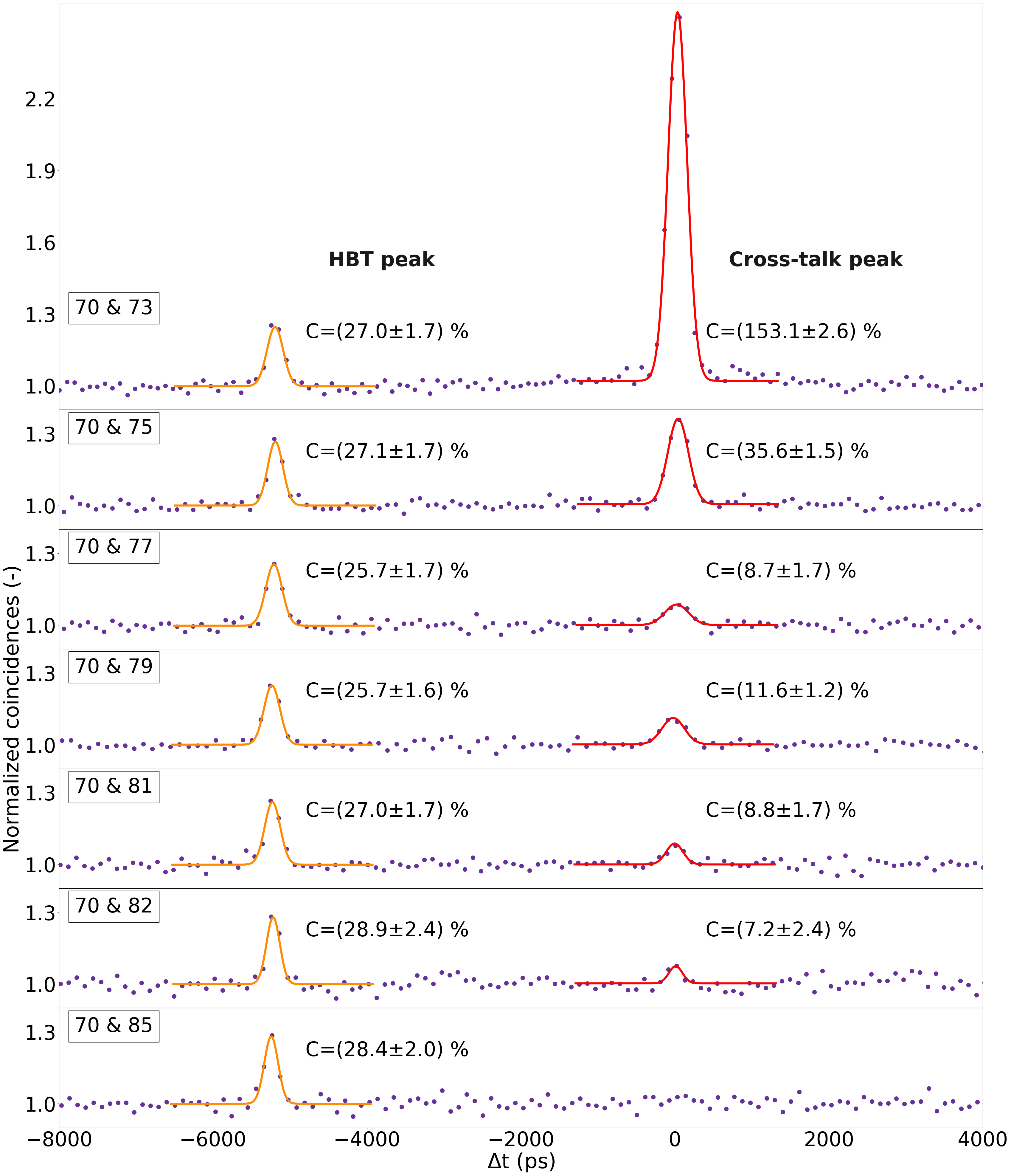}
		\caption{Joint plot of CT and HBT measurements for seven different pairs of pixels. One of the beams was focused on pixel 70 throughout all measurements, while the other one was gradually shifted from pixel 73 to 82. The contrasts of the CT and HBT peaks are shown to the right of each peak. The coincidences are normalized and fitted using the Gaussian function. As the distance between the pixels increases, the CT peak diminishes and eventually fully disappears for the last pixel pair, while the HBT peak stays unaffected. The average distance between the HBT and CT peaks caused by the prolongation of one of the beamsplitter's arms via an additional 1 m fiber is $\Delta t = (5227 \pm 10)$ ps, and all measured values fall within the $2\sigma$ interval.}
		\label{fig:ct_vs_hbt}
	\end{figure}
	
\begin{figure}[ht]
		\centering 
		\adjustbox{valign=c}{\includegraphics[width=.47\textwidth]{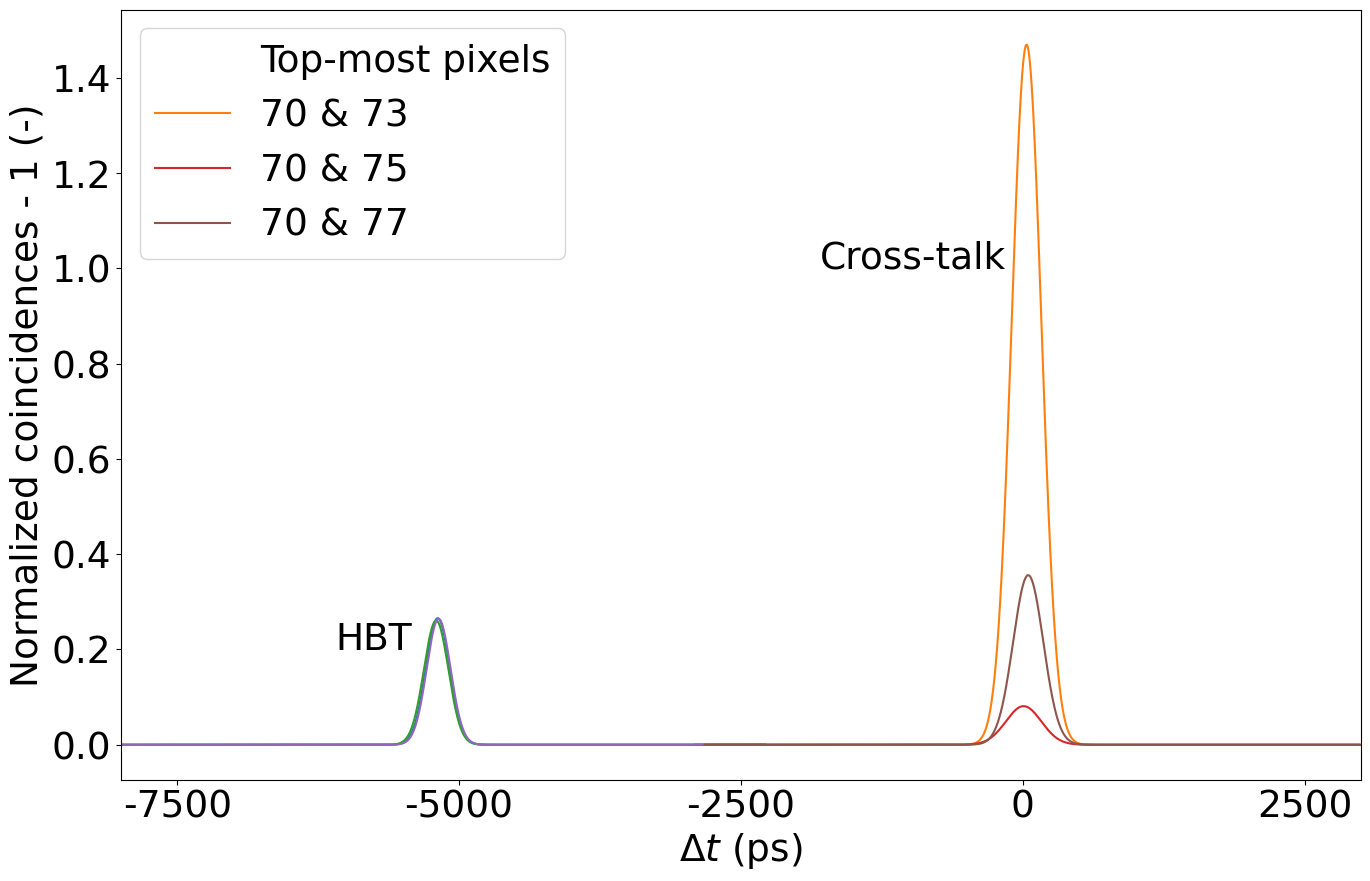}}
		\qquad
		\adjustbox{valign=c}{\includegraphics[width=.47\textwidth]{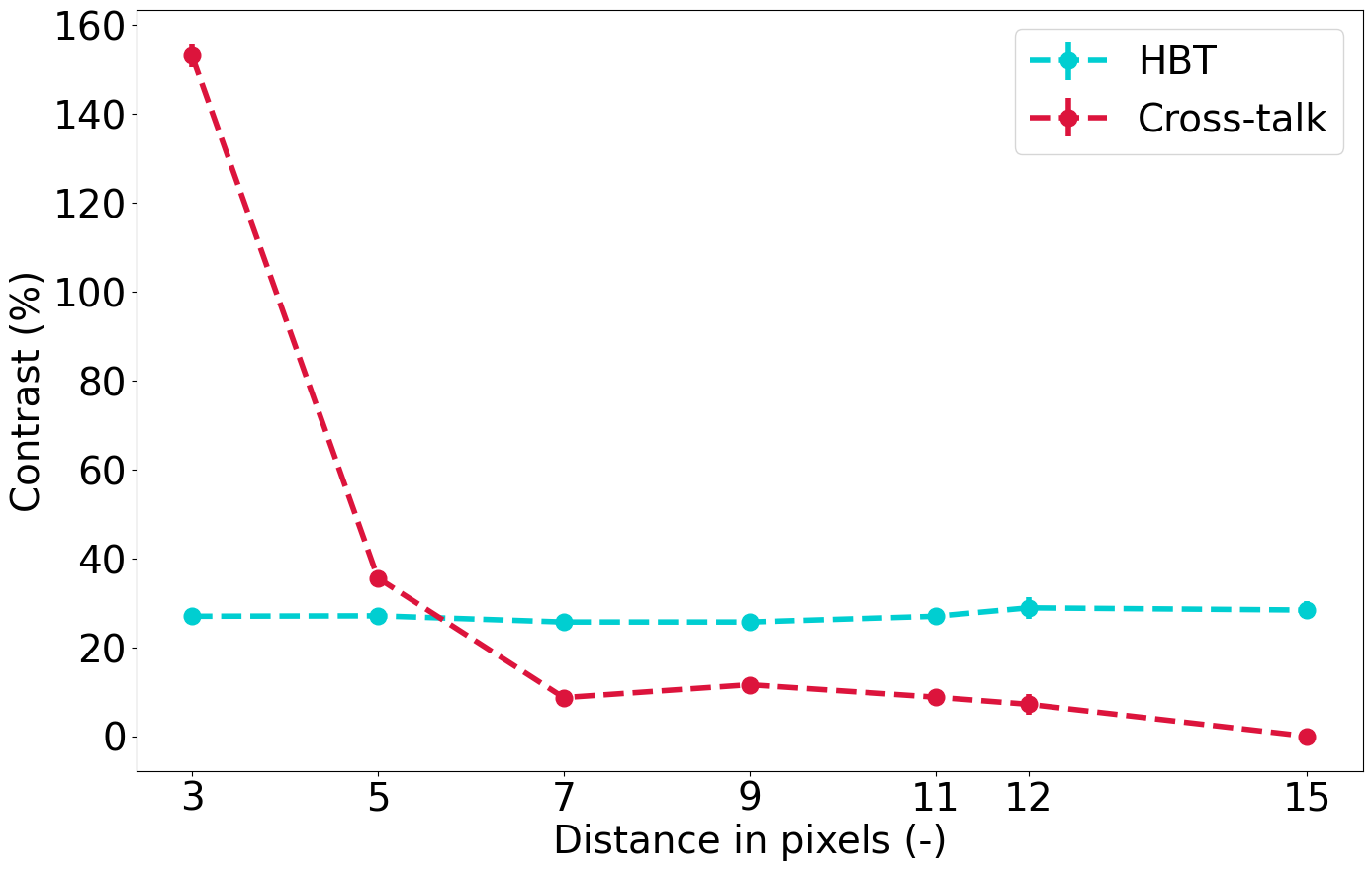}}
		\caption{\label{fig:CT_HBT_Contrast}Left: Joint plot of fits of the CT and HBT peaks for three closest pixel pairs, emphasizing the reduction in the height of the CT peak and the stability of the HBT peak, corresponding to the dependence of their probabilities on distance between the two pixels of interest. While the CT probability decreases with the separation distance of the illuminated pixels, the HBT probability remains constant due to the unchanging conditions for the HBT effect thanks to the single-mode fiber beamsplitter. For each pair, the measured coincidences were normalized by their median value, and the background was subtracted. The CT peaks are aligned to the $\Delta \textit{t}$ = 0. Right: CT and HBT contrasts as a function of the separation distance in pixels. While the HBT contrast mostly remains stable, a sharp decrease in the CT contrast is observed.}
	\end{figure}

	Figure \ref{fig:ct_vs_hbt} shows the results of the joint CT and HBT measurements. All peaks in each data set are fitted using the Gaussian function to extract the peak position and to calculate the contrast. A sharp decrease in the height of the CT peak can be observed as the distance between illuminated pixels increases. This corresponds to the decrease in CT probability with the increasing distance between the pixels at which the two beams are focused. The most noticeable drop is visible for the first three pixel pairs, when one beam was focused on pixel 70 and the other at pixels 73, 75, and 77 in three different measurements. The contrast of the first CT peak, where illuminated spots are separated by a distance of only three pixels, reaches $(148.1 \pm 3.5)\%$. Increasing the separation distance by only four more pixels results in a contrast $(9.2 \pm 1.5)\%$. Shifting the second beam by four pixels thus led to a 16-fold contrast reduction. This can be seen in the left panel in Figure \ref{fig:CT_HBT_Contrast}, where all three CT peaks, as well as the HBT peaks from the three aforementioned pixel pairs, are combined in a single plot. For larger distances, the decrease in the average CT probability is very gradual. For the pixels 70 and 85, the contrast of the CT peak could not be measured as the probability of CT is very low at this distance. In the plot, there is no peak visible, and it was impossible to distinguish between the signal data and the background. The change in contrast of both CT and HBT peaks with the increasing distance between two pixels is shown for all 7 pairs of pixels in the right panel in the right panel in Figure \ref{fig:CT_HBT_Contrast}. 
	
	In comparison, the HBT peak remains stable. Although the HBT effect does depend on the spatial mode of the two photons and is expected to diminish with the increasing distance between the two detectors, we secure the same spatial mode of both HBT photons with the use of a single-mode beamsplitter, ensuring we see the same peak for every pair of illuminated pixels. The average contrast for HBT is $(27.1 \pm 0.7) \% $, and all measured values fall within the $3\sigma$ interval.
	
	\subsection{Using cross-talk for timing offset calibration}
	
\begin{figure}[ht]
		\centering 
		\adjustbox{valign=c}{\includegraphics[width=.47\textwidth]{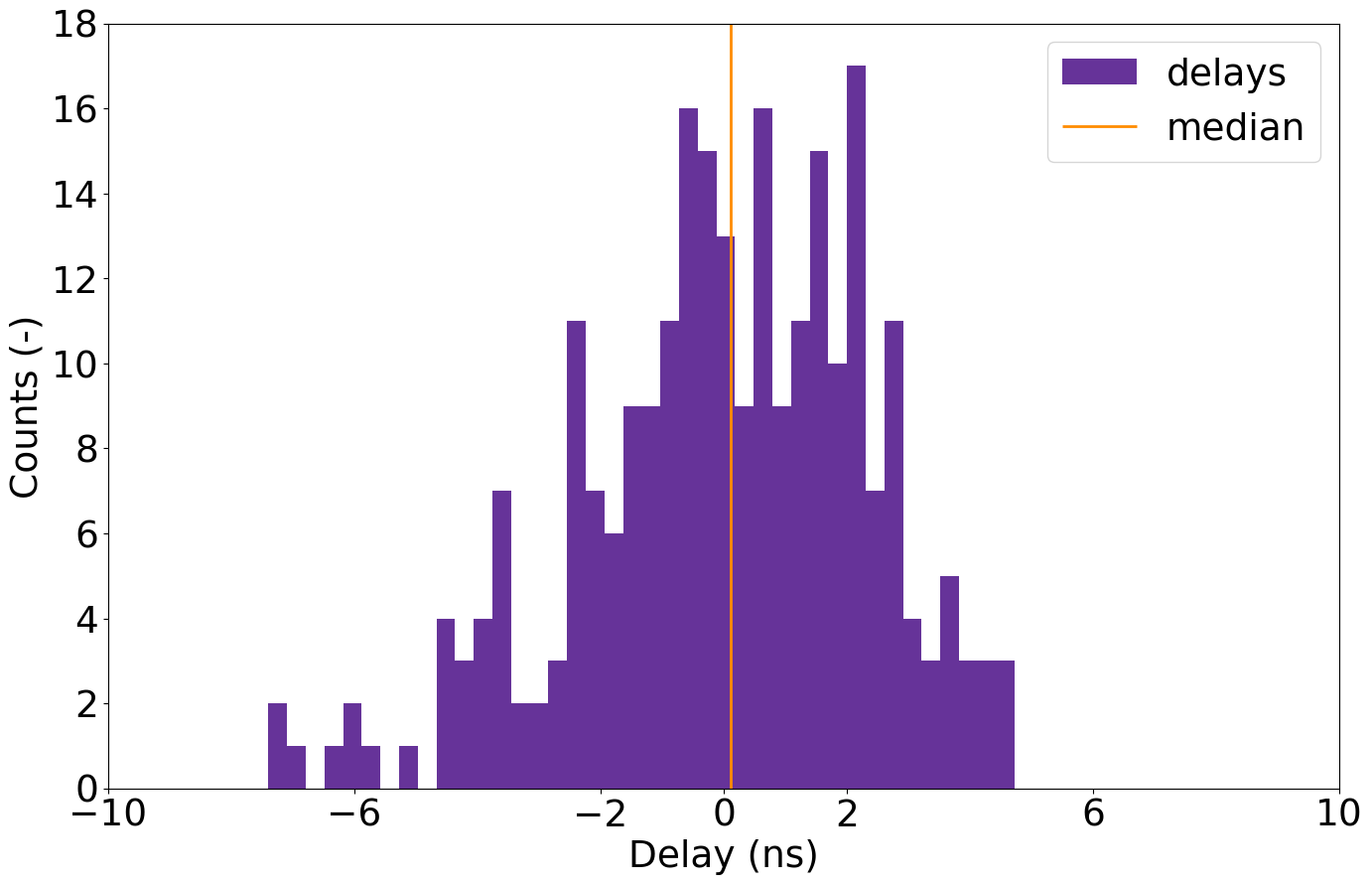}}
		\qquad
		\adjustbox{valign=c}{\includegraphics[width=.47\textwidth]{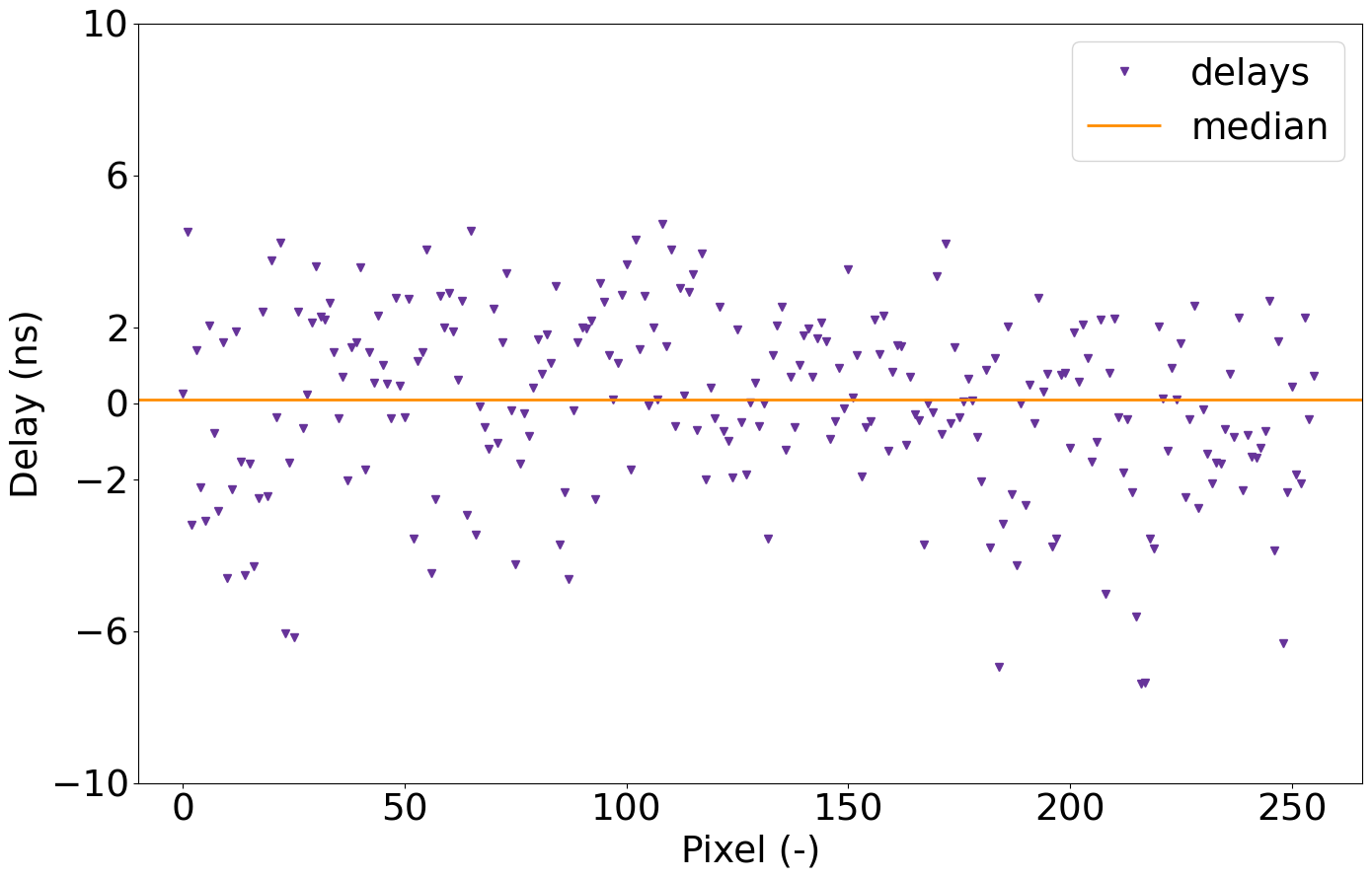}}
		\caption{\label{fig:offset_delays} The calibrated delays. Left: Histogram of the delays with the median. Right: Delay of each pixel with the median. The plots show that individual delays may reach several ns. The median is close to 0 due to the requirement for the average delay equaling 0 set during the calibration, and not exactly 0 due to the widths of the CT peaks used in the calibration.}
	\end{figure}

	Using a data set collected with an LED shining upon the whole sensor, we calculated the individual delays added to each timestamp using the CT between each consequent pair of neighboring pixels. The total data collection time is 20 minutes. The estimated delays can be seen in Fig. \ref{fig:offset_delays}. The individual delays reach several ns, which, for some pairs of pixels, may result in an offset of the coincidence peak on the order of 10 ns.
	
\begin{figure}[ht]
		\centering 
		\adjustbox{valign=c}{\includegraphics[width=.47\textwidth]{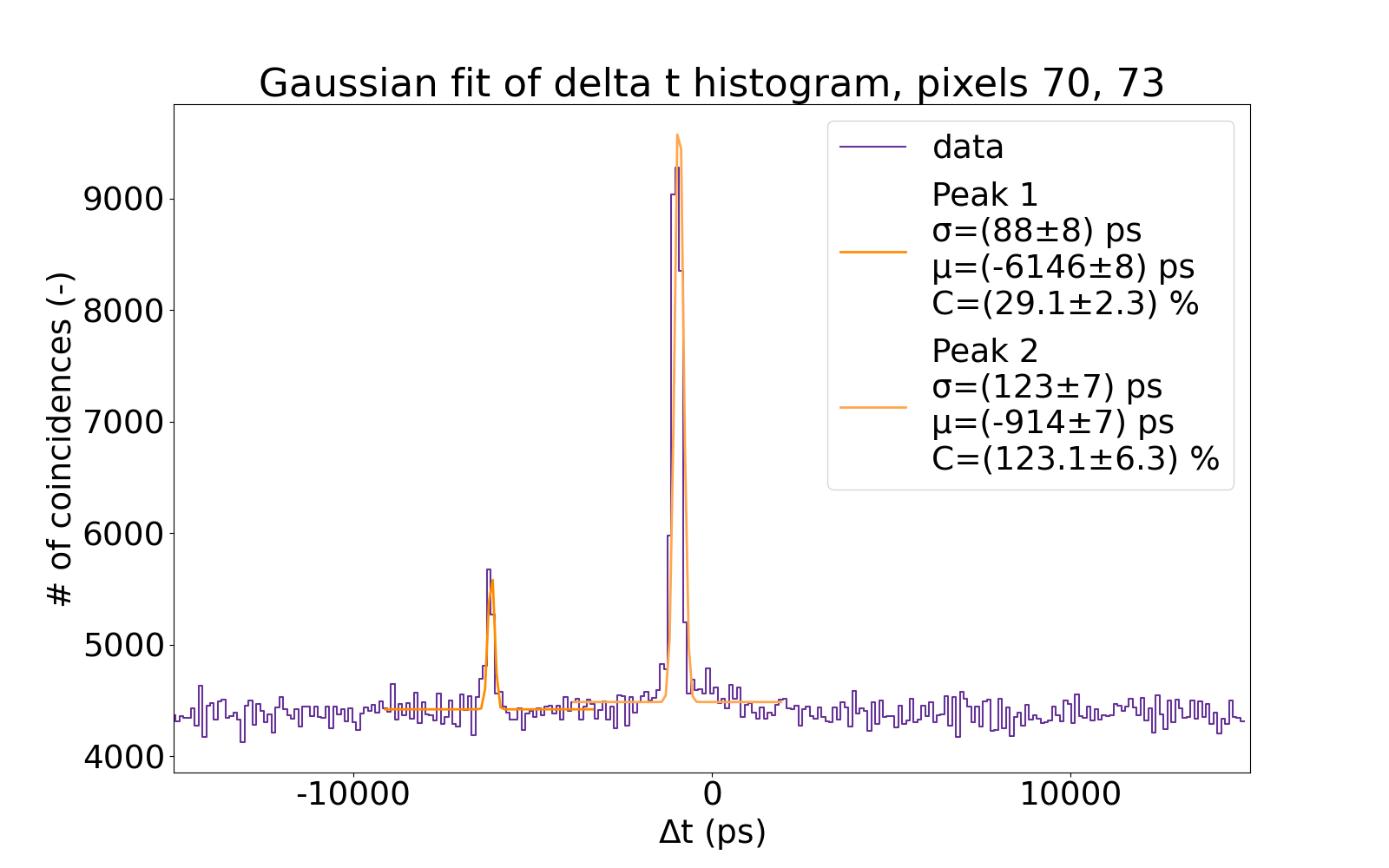}}
		\qquad
		\adjustbox{valign=c}{\includegraphics[width=.47\textwidth]{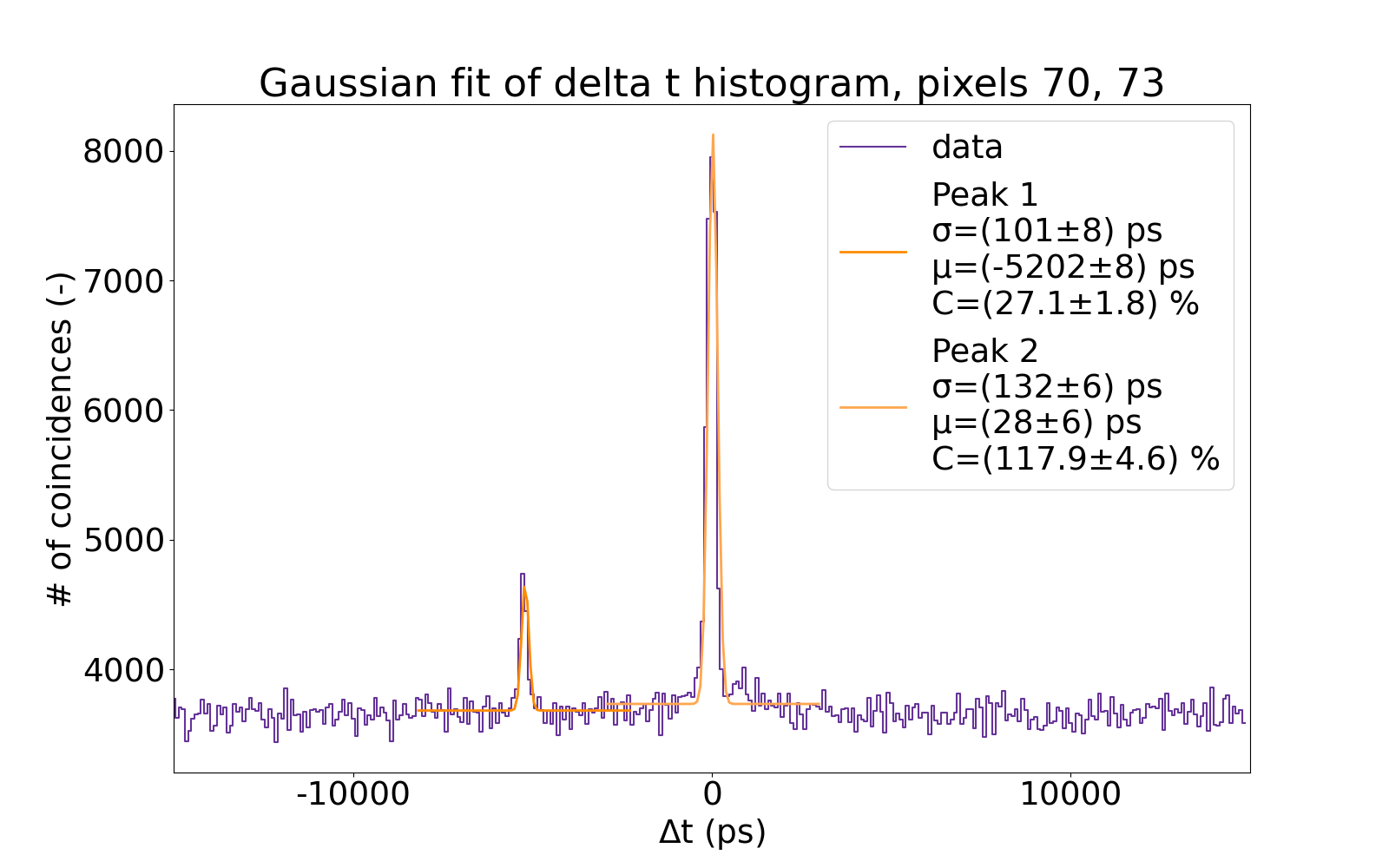}}
		\caption{\label{fig:offset_result} CT and HBT peaks: on the left, before applying the offset calibration, on the right --- after. Both peaks are fitted with the Gaussian function to extract their contrast and position. The HBT peak is shifted relative to the CT one due to an additional 1 m optical fiber inserted to delay one of the two signal photons. After applying the calibration, the CT peak is shifted close to $\Delta t=0$. The calibration is not perfect due to the intrinsic width of the CT peaks that are used for calculating the delays.}
	\end{figure}

	Applying the offset calibration to a data set where, for example, both the HBT and the CT peaks are seen will result in shifting of the two, see Fig. \ref{fig:offset_result}: the CT will appear near the $\Delta t=0$ ns position, while the HBT --- if at least one of the two photons is delayed --- will be shifted according to the delay. The CT peak is not shifted exactly to 0 because of the lack of precision due to the intrinsic width of each CT peak, which is on the order of 100 ps. Nonetheless, this approach is precise enough and can help in discerning between the HBT and CT peaks and can allow for setting bounds for the position of the peak for the Gaussian function to reach a more precise fit. This calibration procedure can be done for each new LinoSPAD2 daughterboard-motherboard combination at the stage of characterizing the CT and DCR, using the CT data for calibrating the offsets. It is important to note that this procedure should be done twice: once for each firmware version, as they differ in how the individual pixels are connected to the corresponding TDCs inside the FPGA.
	
	\section{Conclusion}
	
	In this work, we characterized the DCR and CT effects in a linear array of SPADs, and investigated how CT affects the two-photon interference measurements. We found that the DCR rises with temperature, showing an increase of 30\% for a $3^{\circ}\mathrm{C}$ rise in temperature in close vicinity to the sensor. However, this effect does not noticeably impact the HBT effect measurements since the DCR remains much lower than the photon rate we typically achieve with sources of thermal light, e.g., noble gas lamps or LEDs. Meanwhile, the CT probability appears to be independent of the temperature of the sensor, at least in the temperature range of 20--30$^{\circ}\mathrm{C}$, where we commonly operate.
	
	We also characterized how the average CT probability changes with the distance between two studied pixels and how an additional layer of material --- in our case, a layer of microlenses on top of the array of SPADs --- affects this. We have measured the average CT probability of 0.12\% for the immediate neighbor, which follows a downward trend with increasing separation. In the presence of the microlenses, the CT probability was found to be one to two orders higher at every distance, peaking at $10^{-2}$\% for the third neighbor and plateauing at about $10^{-3}\%$ between sixth and eleventh neighbors. This can be considered a disadvantage for the sensors with microlenses, as one gets higher PDE at the cost of stronger and farther-reaching CT. 
	
	Next, we performed joint CT and HBT measurements to compare these two effects while varying the distance between illuminated pixels. As expected, the CT peak contrast declined with increasing separation, reflecting the lower CT probability at larger distances. In contrast, the HBT peak contrast remained stable, owing to the use of a single-mode fiber beamsplitter that ensures both photons share the same spatial mode for HBT interference. To distinguish the two peaks, one of the photons can be delayed to shift the HBT peak relative to the CT peak. This approach becomes especially important in the presence of microlenses, where CT is stronger and can reach a contrast of up to 9\% at a distance of 12 pixels (over 300 $\upmu\mathrm{m}$ in LinoSPAD2).
	
	Finally, we explored leveraging CT to compensate for intrinsic delays in LinoSPAD2 that otherwise introduce offsets in photon coincidence plots. By illuminating the sensor with ambient light only --- i.e., without additional specialized equipment --- we were able to shift the CT peaks to within $\pm50$ ps of $\Delta t = 0$ in the coincidence histogram. Although the finite width of the CT peaks and uncertainty in their positions introduce some errors, the calibration proves sufficiently accurate, narrowing the original $\pm 10$ ns offset down. This improvement is especially beneficial for precisely fitting the HBT peaks in coincidence histograms, thus enabling more accurate calculations of contrast --- a crucial parameter in many applications, including stellar interferometry.
	
	
	
	\acknowledgments 
	
	This research was supported by the Czech Science Foundation (GACR) under Project No. 25-15534M. This work was also supported by the EPFL internal IMAGING project “High-speed multimodal super-resolution microscopy with SPAD arrays”
	
	\bibliography{references} 
	\bibliographystyle{spiebib} 
	
\end{document}